\documentclass[preprint]{elsarticle}
\usepackage{amsmath,amssymb}
\usepackage{booktabs}
\usepackage{threeparttable}
\usepackage{dcolumn}
\newcolumntype{d}[1]{D{.}{.}{#1}}
\usepackage{epstopdf}
\usepackage[figuresright]{rotating}
\usepackage{xcolor}
\usepackage{textcomp}
\usepackage{geometry}

\journal{Materials \rm{\&} Design}
\biboptions{sort&compress}

\begin{document}
\begin{frontmatter}

\title{Understanding the mechanical properties of reduced activation steels}

\author[DAL,KTH]{Xiaojie Li}
\author[KTH]{Xiaoqing Li\corref{cor}}
\ead{xiaoqli@kth.se}
\author[KTH]{Stephan Sch\"onecker\corref{cor}}
\ead{stesch@kth.se}
\author[CZ]{Ruihuan Li}
\author[DAL]{Jijun Zhao\corref{cor}}
\ead{zhaojj@dlut.edu.cn}
\author[KTH,UU,WI]{Levente Vitos}
\cortext[cor]{Corresponding author}
\address[DAL]{Key Laboratory of Materials Modification by Laser, Electron,
and Ion Beams (Dalian University of Technology), Ministry of Education, Dalian 116024, China}
\address[KTH]{Applied Materials Physics, Department of Materials Science and Engineering,
KTH - Royal Institute of Technology, Stockholm SE-10044, Sweden}
\address[CZ]{Institute of Mold Technology, Changzhou Vocational Institute of Mechatronic Technology, Changzhou 213164, China}
\address[UU]{Department of Physics and Astronomy, Division of Materials Theory,
Box 516, SE-75120 Uppsala, Sweden}
\address[WI]{Research Institute for Solid State Physics and Optics, P.O. Box 49,
H-1525 Budapest, Hungary}

\begin{abstract}
Reduced activation ferritic/martensitic (RAFM) steels are structural
materials with potential application in Generation-IV fission and
fusion reactors. We use density-functional
theory to scrutinize the micro-mechanical properties of the main
alloy phases of three RAFM steels based on the body-centered cubic FeCrWVMn solid solution. We  assess the lattice parameters and elastic properties
of ferromagnetic
$\alpha$-Fe and Fe$_{91}$Cr$_{9}$, which are the main building
blocks of the RAFM steels, and present a detailed
analysis of the calculated alloying effects of V, Cr, Mn, and W on
the mechanical properties of Fe$_{91}$Cr$_{9}$. The composition
dependence of the elastic parameters is decomposed into electronic
and volumetric contributions and studied for alloying levels that
cover the typical intervals in RAFM steels. A linear
superposition of the individual solute effects on the
properties of Fe$_{91}$Cr$_{9}$ is shown to provide an excellent
approximation for the \emph{ab initio} values obtained for the
 RAFM steels. The intrinsic ductility is evaluated through Rice's phenomenological theory using the surface and unstable stacking fault energies, and the predictions are contrasted with those obtained by empirical criteria. Alloying with V or W is found to enhance the ductility,
whereas additional Cr or Mn turns the RAFM base alloys more brittle.
\end{abstract}


\end{frontmatter}

\section{Introduction}

The structural materials of the first wall, breeding-blanket and divertor components of future fusion power reactors, such as the International Thermonuclear Experimental
Reactor (ITER) and the Demonstration
Power Plant (DEMO), will be exposed to plasma particles and 14\,MeV neutron irradiation. As high-energy neutrons will cause displacement damage and nuclear transmutation reactions in these components,  
the development of reduced-activation materials (i.e., with a minimum amount of elements that would result in long-lived radioactive isotopes) for structural applications capable of withstanding a high neutron fluence is one of the most critical challenges in fusion technology research.
For instance, the estimated key irradiation parameters of the first wall in DEMO with a fusion power of 2-2.5\,GW in operation include a neutron wall loading of $<2$\,MW/m$^{2}$ and a neutron fluence of 5-8\,MW-y/m$^2$, which would amount to an accumulated dose of 25-30\,dpa per year in steels~\mbox{\cite{moslang2005towards}}.

Reduced activation ferritic/martensitic (RAFM) steels based on low-activation elements (e.g., Fe, V, Cr, Mn, Ta, W, Si, C) are currently one of the most promising structural materials for first wall and breeding-blanket applications in fusion reactors~\mbox{\cite{klueh2001high,baluc2003potentiality,ehrlich1996development,Baluc:2007a}}. 
They were selected for the test blanket module for ITER~\mbox{\cite{Schaaf:2006,Baluc:2007a,wang2014microstructural,shi2015microstructure}} and are considered as a primary candidate structural material for DEMO~\mbox{\cite{hishinuma1998current,de2016wettability}}.
RAFM steels are essentially modifications of the body-centered cubic (bcc), Fe-rich, Fe-Cr
binary alloys and contain minor concentrations of low-activation elements, such as 
manganese to improve the abrasion and
wear resistance as well as tensile properties~\cite{Gunn:1997},
tungsten and vanadium to maintain a low activation level and to
resist irradiation embrittlement~\cite{Tavassoli:2002,Lindau:2001}.
The composition of the main alloying elements lies in the range (wt.\%) Fe-(7.5-12)Cr-(1.0-2.2)W-(0.15-0.25)V-(0.05-0.6)Mn\mbox{~\cite{Baluc:2007b,Huang:2007}}.

Recent experimental progress has mainly focused on the fabrication, manufacturing, mechanical properties (precipitation behavior, fracture toughness, creep, fatigue, and thermal aging), effects of irradiation, and corrosion analysis of RAFM steels\mbox{~\cite{Schaaf:2000,Baluc:2007b,Huang:2013,manugula2016critical,Kimura:2005,li2007mechanical}}.
Irradiation damage on the microstructure and mechanical properties, including irradiation hardening and embrittlement by neutrons and helium, fatigue and creep after irradiation, were intensively investigated\mbox{~\cite{Kimura:2005,Baluc:2007b,Huang:2013,zhao2008effect}}.
RAFM steels turned out to be far superior to austenitic steels in terms of swelling and helium embrittlement resistance. For instance, a swelling rate  of approximately 1\,vol.\% at 575\,K is reached in RAFM steels after a dose of 100\,dpa, whereas the same swelling rate is already reached after 10\,dpa in typical austenitic steel\mbox{~\cite{Huang:2013}}. 
From the technological point of view, the mechanical properties of the structural materials mainly limit the temperature window of operation of fusion reactors. In the case of RAFM steels, this interval is approximately 620-820\,K (possibly higher in oxide dispersion strength variants) owing to irradiation-induced embrittlement effects and a loss in mechanical strength with temperature\mbox{~\cite{lucon2006european,Baluc:2007b}}.

The knowledge of the elastic properties and solute-induced changes in these parameters is important for alloy design since they directly provide information about the mechanical response to various loading conditions as well as mechanical stability. 
Moreover, the elastic coefficients are necessary parameters for a multi-scale modeling approach to the mechanical properties of alloys. 
For instance, interatomic potentials for atomistic simulations of crystal defects are typically validated against elastic constants determined experimentally or predicted from density-functional theory (DFT) calculations. 
The knowledge of the elastic parameters allows estimating the critical stress for twinning nucleation\mbox{~\cite{Kibey:2007}}, dislocation core properties\mbox{~\cite{Hirth:2002}}, and solid solution hardening\mbox{~\cite{Nabarro:1977,Labusch:1972}}, they enter crystal plasticity\mbox{~\cite{Roters:2010}} and phase-field models\mbox{~\cite{Yeddu:2012}}, and were related to ductility\mbox{~\cite{Pugh:1954}}.
Despite the significant experimental efforts for RAFM steels mentioned above, little is known about the effects of solutes on the elastic properties, or the Fe-rich Fe-Cr binary as an approximant to the ferritic phase of RAFM steels. 
Noteworthy, the polycrystalline elastic constants of Fe-Cr were determined by Speich \emph{et
al.}~\cite{Speich:1972}. 
On the theoretical side, the single and polycrystalline elastic
properties of the Fe-Cr solid solutions were assessed by
Korzhavyi \emph{et al.}~\cite{Korzhavyi:2009} and Zhang \emph{et
al.}~\cite{Zhang:2010b} by means of DFT
calculations. 
Recently, Xu \emph{et al.}~\cite{Xu:2014} studied the
elastic properties of the Fe-Cr-W solid solution with Cr content in
the range of 7.8-10 wt.\% (8.4-10.7 at.\%) and W content in the
range of 1-2 wt.\% (0.3-0.6 at.\%). Apart from these theoretical
investigations on the Fe-Cr binary and Fe-Cr-W ternary alloys in a
narrow tungsten concentration range, little is known about the
alloying effects of other substitutional solutes commonly employed
in RAFM steels, such as V and Mn. Furthermore, most of the above
studies were restricted to the elastic moduli without going beyond
 simple empirical correlations between these bulk elastic parameters and
ductile/brittle behavior.

In this paper, we use DFT to scrutinize the micro-mechanical properties of the main alloy phases of RAFM steels.
We selected the CLAM/CLF-1~\cite{Huang:2007,Xia:2010}, F82H~\cite{Jitsukawa:2002}, and EUROFER97~\cite{Schaaf:2003} grades with chemical compositions given in Table~\ref{table:1} (main alloying elements) and concentrate on the ferritic (bcc) phase.
Other substitutional or interstitial elements in low concentrations, such as C, Ta, S, P,
O, and N, may be present in real RAFM materials (C and N stabilize the martensitic phase)~\cite{Baluc:2007b}, but they are not considered here.
CLAM/CLF-1, F82H, and EUROFER97 were chosen since they have achieved great technology maturity and are under extensive investigation on an industrial scale\mbox{~\cite{Baluc:2007b,Huang:2013}}.
Although RAFM steels in power reactors will be subject to ambient temperature and significantly above, we here focus on the low-temperature region, i.e., all DFT results obtained in this investigation are relevant for zero Kelvin. 
The reason is that experimental low, ambient, and high-temperature elastic properties of non-irradiated RAFM steels are currently missing except for the elastic moduli of F82H\mbox{~\cite{Tavassoli:2002}} and CLAM/CLF-1\mbox{~\cite{Xu:2014}} determined at room temperature.
We are concerned with the lattice parameter, single-crystal and polycrystalline elastic properties, and intrinsic ductility of CLAM/CLF-1, F82H, and EUROFER97.
We use the computed elastic parameters and empirical relationships\mbox{~\cite{Pugh:1954,Pettifor:1992}} to examine the intrinsic ductility of these grade. We compare these predictions with those derived from the phenomenological model of Rice\mbox{~\cite{Rice:1992}} employing the surface and the unstable stacking fault (USF) energies.
Choosing the Fe$_{91}$Cr$_9$ binary as base alloy, since its chemical composition is close to those of CLAM/CLF-1, F82H, and EUROFER97 (Table~\ref{table:1}), we deepen the investigation of the elastic parameters by considering binary Fe$_{91-c}$Cr$_{9+c}$ (additional Cr) and ternary Fe$_{91-c}$Cr$_{9}M_c$, $M$=W, V, or Mn, random solid solutions. 
We examine concentration intervals $c$ larger than those of typical RAFM compositions (Table\mbox{~\ref{table:1}}) in order to establish robust alloying coefficients for Cr, W, V, or Mn. We quantify the solute induced volumetric and electronic contributions to the changes in the single-crystal elastic constants for each solute. We propose and validate a linear superposition rule, which allows predicting the single-crystal and polycrystalline elastic properties of quinary RAFM steels based on the alloying coefficients of Cr, W, V, and Mn (Cr relative to 9\,at.\%) solutes in Fe$_{91}$Cr$_9$.

The remainder of the paper is organized as follows.
A brief account of the methodological and computational details is given in \mbox{Sec.~\ref{sec:methodology}}.
Section~\ref{sec:res} presents the main results. We start by assessing our DFT methodology for
the lattice parameter, elastic properties, surface and the unstable stacking fault energies of bcc ferromagnetic
$\alpha$-Fe and Fe$_{91}$Cr$_{9}$.
We then examine the same parameters for CLAM/CLF-1, F82H, and EUROFER97.
In Sec.~\ref{sec:discussion}, we analyze and discuss the elastic properties of the Fe$_{91-c}$Cr$_{9+c}$ and Fe$_{91-c}$Cr$_{9}M_c$, $M$=W, V, or Mn, solid solutions, and discuss the alloying effects on the intrinsic ductility. 
The major results are summarized in Sec.~\ref{sec:conclusion}.

\begin{table*}
\caption{\label{table:1}Chemical compositions for the bcc, quinary
Fe$_{100-x-y-z-u}$Cr$_{x}$W$_{y}$V$_{z}$Mn$_{u}$ alloys employed to model the specific chemical compositions of the
selected RAFM steels~\cite{Baluc:2007b,Huang:2013}. Data are listed both in atomic and weight percent.}
\begin{tabular}{lll}
\toprule
Grade & Atomic percent (\%) & Weight percent (\%) \\
\midrule
CLAM/CLF-1 &  Fe$_{88.93}$Cr$_{9.63}$W$_{0.45}$V$_{0.53}$Mn$_{0.46}$ & Fe$_{88.85}$Cr$_{9.0}$W$_{1.5}$V$_{0.2}$Mn$_{0.45}$ \\
F82H & Fe$_{90.77}$Cr$_{8.02}$W$_{0.60}$V$_{0.40}$Mn$_{0.21}$ & Fe$_{90.22}$Cr$_{7.46}$W$_{1.96}$V$_{0.15}$Mn$_{0.21}$ \\
EUROFER97 & Fe$_{89.07}$Cr$_{9.49}$W$_{0.33}$V$_{0.66}$Mn$_{0.45}$ & Fe$_{89.3}$Cr$_{8.9}$W$_{1.1}$V$_{0.25}$Mn$_{0.45}$ \\
\bottomrule
\end{tabular}
\end{table*}

\section{\label{sec:methodology}Computational method}

\subsection{Lattice parameter and elastic properties}

The equilibrium lattice constant and bulk modulus were derived from
the equation of state determined by fitting a Morse-type
function~\cite{Moruzzi:1988} to the total energy data
evaluated at seven different volumes.
Using instead the Murnaghan~\mbox{\cite{murnaghan1944compressibility}}, Birch-Murnaghan~\mbox{\cite{birch1947finite}}, Vinet~\mbox{\cite{vinet1987compressibility}}, or Poirier-Tarantola~\mbox{\cite{Poirier:1998}} equations of state gave very similar results (scatter of lattice constant $<10^{-3}$\,\AA, scatter of the bulk modulus $<1\,$GPa) and resulted in identical compositional trends.

For a cubic crystal, there are three
independent second-order elastic constants, $C_{11}$, $C_{12}$, and
$C_{44}$. Here, $C_{11}$ and $C_{12}$ were obtained through the
tetragonal shear modulus $C^\prime= (C_{11}-C_{12})/2$ and the bulk
modulus $B= (C_{11} + 2C_{12})/3$. The following volume-conserving
orthorhombic ($D_\text{o}$) and monoclinic ($D_\text{m}$) strains
were used to calculate $C^\prime$ and $C_{44}$, respectively,
\begin{equation}
D_{\text{o}} = \left( \begin{array}{rrr} 1+\delta_\text{o} & 0 & 0 \\  0 & 1-\delta_{\text{o}} &  0 \\ 0 & 0 & \frac{1}{1-\delta^2_{\text{o}}} \end{array}\right)
\end{equation}
and
\begin{equation}
D_{\text{m}} = \left( \begin{array}{rrr} 1 & \delta_\text{m} & 0 \\ \delta_\text{m} & 1 &  0 \\ 0 & 0 & \frac{1}{1-\delta^2_{\text{m}}}  \end{array}\right)  ,
\end{equation}
where $\delta $ denotes the strain parameter. The two deformations
lead to the energy changes $\Delta E(\delta_{\text{o}}) \equiv E(\delta_{\text{o}}) - E(0) =2VC^\prime
\delta^2_{\text{o}} + \mathcal{O}(\delta^4_{\text{o}})$ and $\Delta
E(\delta_{\text{m}}) \equiv E(\delta_{\text{m}}) -  E(0) =2VC_{44} \delta^2_{\text{m}} +
\mathcal{O}(\delta^4_{\text{m}})$. 
The previous equations were fitted to the total energy differences computed for six distortions ($\delta=0.00, 0.01,\ldots, 0.05$) using a least squares method.

The polycrystalline shear modulus $G$ was computed from the
single-crystal elastic constants using the Hill average
method~\cite{Hill:1952,Vitos:2007}, whereas the polycrystalline
Young modulus $E$ and Poisson's ratio $\nu$ are related to $B$ and
$G$~\cite{Vitos:2007} through
\begin{subequations}
    \begin{align}
    E &= \frac{9BG}{3B+G}\qquad \text{ and }\\   	
    \nu &= \frac{3B-2G}{2(3B+G)},\label{eq:polyEnu}    
    \end{align} 
\end{subequations}
respectively. The Hill's method and Eqs.~\eqref{eq:polyEnu} were
also employed to derive the isotropic moduli from experimental
single-crystal data.

\subsection{Planar fault energies}

\begin{figure}[tbh]
    \resizebox{\columnwidth}{!}{\includegraphics[clip]{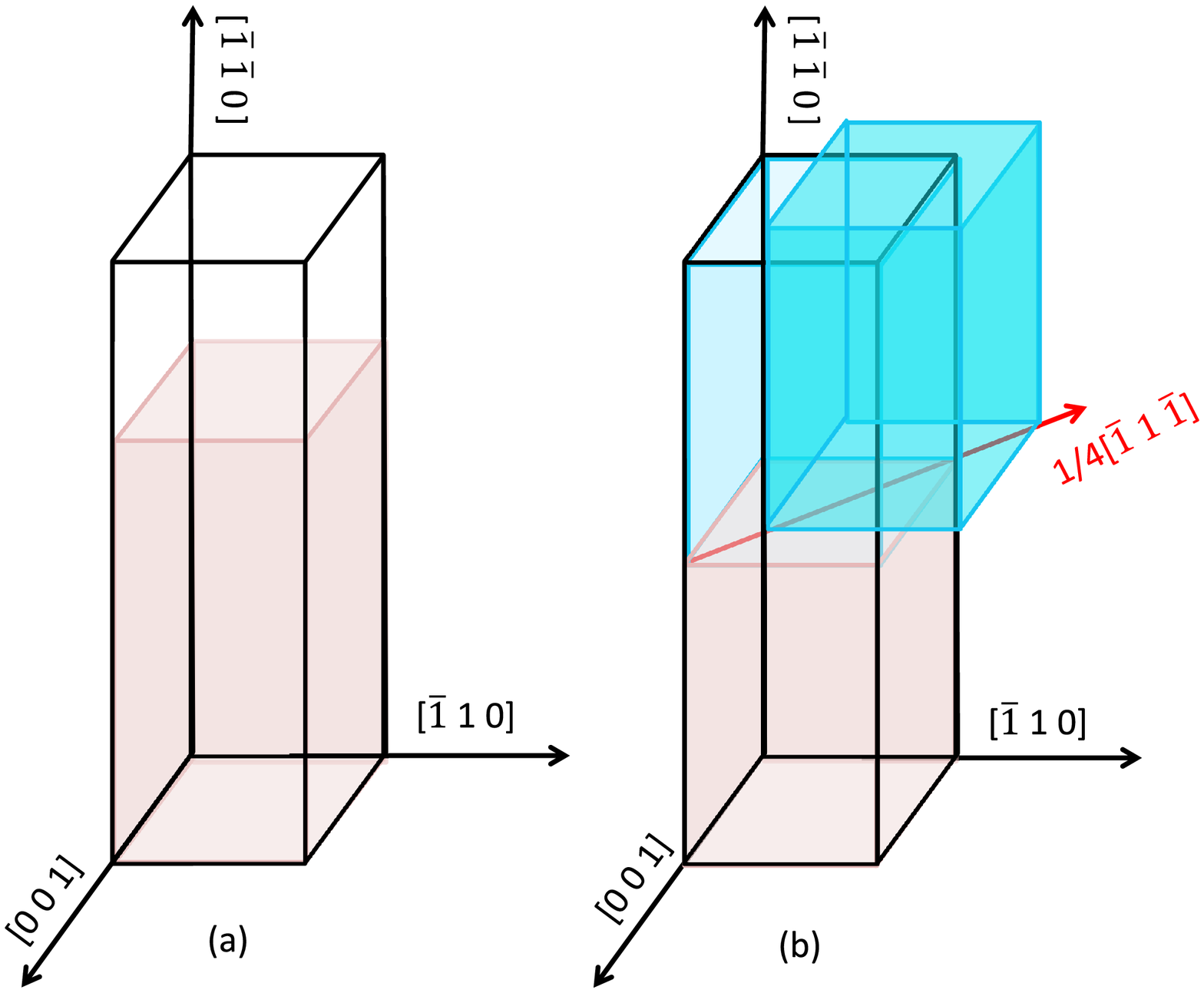}}
    \caption{\label{fig:surfacemodel}
    Schematic simulation boxes (subject to periodic boundary conditions)
    used to compute the surface energy of the $\{110\}$ surface facets
    (panel a) and the USF energy of the $\{110\}\langle 111\rangle$ slip
    system (panel b). The main crystallographic directions are indicated.
    In (a) the colored part and the white part of the box represent the
    material and the vacuum, respectively. In (b) the upper part is shifted
    against the lower one by one half of the Burgers vector ($1/4 [\bar{1}1\bar{1}]$)
    indicating the unstable stacking fault (dark blue) and the original position (light blue).}
\end{figure}

The surface energy $\gamma_{s}$ and the unstable stacking fault
(USF) energy $\gamma_{u}$ of the bcc $\{110\}$ planes are employed to
estimate the alloying effect on the intrinsic ductility within the model reported by Rice~\cite{Rice:1992}; see  Sec.~\ref{sec:ductility}
We considered chemically homogeneous alloys in the determination of both planar fault energies. This is motivated as follows.
The Rice model assumes the presence of a sharp crack tip in an otherwise defect-free material and estimates the competition between crack propagation (characterized by the surface energy) versus crack blunting by dislocation nucleation (characterized by the USF) under an appropriate, externally applied stress.
The dislocation is assumed to be emitted into a chemically homogeneous bulk matrix, and segregation to the newly created crack surface 
is expected to occur (if at all) after the surface cleaved, on a much longer timescale than the one governing the propagation of the crack.
Thus, we did not consider surface segregation and constructed chemically homogeneous simulation cells in the determination of $\gamma_{s}$ and $\gamma_{u}$ as detailed in the following.

We recall that for
bcc Fe, the close-packed $\{110\}$ surface facets possess the lowest
surface energy~\cite{Blonski:2007}. 
The schematic surface model for the
$(110)$ plane is shown in Fig.~\ref{fig:surfacemodel}(a). Surface
energies are usually computed from slab models and several
procedures have been proposed to yield converged
$\gamma_{\text{s}}$'s with respect to the slab
size~\cite{Boettger:1994,Fiorentini:1996}. Here, the surface energy
was calculated from the total energy of surface slabs using the
method reported by Fiorentini and Methfessel~\cite{Fiorentini:1996}.
In this approach, the bulk energy is obtained from a fit of the slab
total energy versus the number of atomic layers in the slab, which
is a linear function of the slab thickness for sufficiently large
slabs (the slope gives the bulk energy). We find that slabs with 7, 9,
and 11 atomic layers decoupled by vacuum of thickness corresponding
to seven bulk interplanar distances yield converged $\{110\}$
surface energies.

Previous theoretical investigations showed that the enhanced surface
magnetism in bcc Fe and Fe-rich Fe-Cr alloy suppresses large
relaxations at the (110) surface facet, and that the actual
relaxation of the surface geometry lowers the surface energy only in
the order of
1\,\%~\cite{Schoenecker:2013,Punkkinen:2011,Punkkinen:2011b,Blonski:2007}.
Here, the relaxation of the surface geometry was determined for the
surface slabs with 11 atomic layers and the impact of relaxation was
evaluated with respect to perfectly truncated bulk crystals with the
same thickness. For pure Fe, the spacing between the surface and the
subsurface layer was found to decrease by approximately 1.3\,\%. For
the study of the surface energy of Fe$_{91}$Cr$_{9}$ and the present
RAFM steels, the maximum solute concentration is about 11\,\% and
expected to result in a small additional relaxation effect. Indeed,
we found for these alloys that surface layer relaxed by 1.4\,\% to
1.5\,\%, similar to pure Fe.

Slip in bcc metals commonly occurs in the $\{110\}$ planes along the
$\langle 111 \rangle$ directions with Burgers vector $\langle 1/2,
1/2, 1/2 \rangle $. The USF energy was thus determined for this slip
system adopting the computational scheme detailed in
Ref.~\cite{Wang:2015} and schematically shown in
Fig.~\ref{fig:surfacemodel}(b). Accordingly, the total energies for
supercells modeling the USF configuration (one
half of the crystal shifted above the other half by one half the
Burgers vector) were computed for various supercell sizes, and the
bulk energy was obtained from a fit of the supercell total energy
versus slab thickness. As for the surface energies, this relation is
linear for sufficiently large supercells, and cells with 16, 20,
and 24 atomic layers were used in the fitting. The interplanar
relaxation at the fault plane was taken into account and the relaxed
USF energies were evaluated for the supercell with 16 layers. We
found that the first interlayer spacing at the USF increases by
4.9\,\% for pure bcc Fe. The relaxations obtained for
Fe$_{91}$Cr$_{9}$, CLAM/CLF-1, F82H, and EUROFER97 range from
4.6\,\% to 4.8\,\%.

\subsection{Electronic structure calculations}

The present results are based on DFT\mbox{~\cite{hohenberg1964inhomogeneous}} performed using the
all-electron exact muffin-tin orbitals (EMTO) total energy
method~\cite{Vitos:2001a,Vitos:2000,Andersen:1994}. The EMTO method
is an improved screened Korringa-Kohn-Rostoker method, in which the
full potential is represented by overlapping muffin-tin potential
spheres. The present overlapping potential spheres describe the
exact potential more than 10-15\,\% more accurately than conventional muffin-tin or
non-overlapping approaches~\cite{Vitos:2001a,Andersen:1998}. 
The radii of the muffin-tin spheres were set to be identical to the Wigner-Seitz radius, since this choice minimizes the muffin-tin discontinuity and ensures a potential sphere overlap error much less than 0.1\,mRy~\mbox{\cite{Vitos:2007}}.

The
self-consistent calculation were carried out using the local-density
approximation to describe the exchange-correlation
interactions~\cite{Perdew:1992}, whereas we used the
Perdew-Burke-Ernzerhof (PBE) functional~\cite{Perdew:1996} within
the generalized-gradient approximation for the total energy. The
chemical disorder was treated by the coherent-potential
approximation (CPA)~\cite{Gyorffy:1972,Taga:2005,Vitos:2001b}. 

The CPA solves the electronic structure problem for random alloys by determining the Green function for an effective medium. The effective medium possess the symmetry properties of the underlying lattice, i.e., the bcc Bravais lattice (space group $Im\bar{3}m$) in this investigation.
Thus, the primitive bcc unit cell suffices to model the bulk properties of all the present alloys, and the simulation cells for the planar defects are derived from supercells thereof (for details, see previous section).
Since the CPA is a single-site approximation, short-range order effects
and local-lattice relaxation were not considered.  
The accuracy of the EMTO-CPA method for the elastic properties and planar fault energies of multi-component alloy systems was demonstrated in previous works~\mbox{\cite{tian2017alloying,Li:2014,li2016ab,Huang:2006,Li:2012,Ropo:2005,Wang:2015,skripnyak2018ab,zhang2016first,huang2016mechanism}}.

\subsection{Numerical details}

In the present electronic structure and total energy calculations,
the one-electron equations were solved with the soft-core scheme
using the scalar-re\-la\-ti\-vis\-tic approximation, and the total energy
was calculated using the full charge-density technique\mbox{~\cite{Vitos:2007}}. 
In the basis
set, we included $\textit{s}$, $\textit{p}$, $\textit{d}$ and
$\textit{f}$ orbitals. 
The electrostatic correction to the single-site
CPA was described using the screened impurity model~\cite{korzhavyi1995madelung} with universal screening parameter 0.607 previously determined for random alloys~\cite{ruban2002screened}.
In the elastic constant calculations and the determination of the bulk modulus, 
the Brillouin zones were sampled by a $\Gamma$-centered, $35\times 35\times 35$ $k$-point mesh uniformly distributed over the Brillouin zone. This settings ensures a convergence of 
these parameters at the level of $<0.3$\,GPa with respect to a denser, $37\times 37\times 37$ $k$-point mesh.
For the determination of the planar fault
energies, $13\times 33\times 1$ $k$-points were found to yield
converged surface energies for the bcc (110) surface facet, whereas
a $13\times 33\times 2$ partitioning was used for the USF energy
calculation. 
All calculations were performed for collinear magnetic configurations.

\section{\label{sec:res}Results}

\subsection{\label{sec:assessment}Lattice parameter and elastic properties of ferromagnetic $\alpha$-Fe and Fe$_{91}$Cr$_{9}$}

In order to analyze the alloying effects of the considered solutes
(V, Cr, Mn, and W) on elastic properties, it is necessary to know
these parameters for the host lattice. Therefore, first we
briefly discuss the lattice constant and the elastic properties of
$\alpha$-Fe (ferromagnetic, bcc phase). Second, we consider the same
quantities for the bcc Fe$_{91}$Cr$_{9}$ matrix, for which the Cr
content is close to that of the considered RAFM steels.

\begin{sidewaystable*}

\caption{\label{table:2}Theoretical (0\,K) and experimental (temperature as stated in the footnotes) lattice parameter $a$, single-crystal elastic constants $C_{11}$, $C_{12}$, and $C_{44}$, shear elastic constant $C^\prime$, Cauchy pressure ($C_{12}-C_{44}$), and Zener anisotropy ratio $C_{44}/C'$ for bcc Fe, Fe$ _{91} $Cr$ _{9}, $CLAM/CLF-1, F82H, and EUROFER97. PAW and FPLMTO stand for projector augmented wave method and full-potential linear-muffin-tin-orbital method, respectively.}
\begin{threeparttable}
\begin{tabular}{*{9}{l}}
\toprule
Material & Method & $a$ (\AA) & $C_{11}$ (GPa) & $C_{12}$ (GPa) & $C_{44}$ (GPa) &  $C^\prime$ (GPa) & $C_{12}-C_{44}$ (GPa) & $C_{44}/C'$\\
\midrule
Fe & This work & 2.838 & 292.7 & 137.8 & 106.5 & 77.4 & 31.3 & 1.37 \\
& EMTO~\cite{Zhang:2010b}  & 2.837 & 297.8 & 141.9 & 106.7 & 77.9 & 35.6 & 1.37 \\
& PAW~\cite{liu2005structure} & 2.827 &  286 & 148 & 101 & 69 & 47 & 1.46  \\
& PAW~\cite{fellinger2017ab} & 2.832 & 278 & 148 & 98 & 65 & 50 & 1.51 \\
& FPLMTO~\cite{Sha:2006} & 2.812 & 303 & 150 & 126 & 76.5 & 44 & 1.60 \\
& Expt.~\cite{Rayne:1961}\tnote{a}  & 2.866 & $243.1\pm 0.8$ & $138.1 \pm 0.4$ & $121.9 \pm 0.4$ & $52.5$ & $ 16.2 \pm 0.8$ & $2.32 \pm 0.01$ \\
& Expt.~\cite{Adams:2006}\tnote{b} &  - & 239.6 & 135.8 & 120.8 & 51.9 & 15.0 & 2.33 \\
\hline
Fe$_{91}$Cr$_{9}$ & This work & 2.849  & 282.9 & 125.3 & 119.3 & 78.8 & 6.0 & 1.51 \\
Fe$_{90}$Cr$_{10}$&  EMTO~\cite{Zhang:2010b} & 2.847 & 287.8 & 127.9 & 120.7 & 79.9 & 7.2 & 1.51  \\
\hline
CLAM/CLF-1 & This work & 2.851 & 285.4 & 126.1 & 120.4 & 79.7 & 5.6 & 1.51 \\
F82H & This work & 2.852  & 283.2 & 126.2 & 118.6 & 78.5 & 7.6 & 1.51 \\
EUROFER97 &  This work & 2.851 & 285.2 & 125.9 & 120.3 & 79.6 & 5.6 & 1.51\\
\bottomrule
\end{tabular}
\begin{tablenotes}
\item[a] $a$ measured at 293\,K; elastic constants measured at 4.2\,K
\item[b] 0\,K data extrapolated from 3\,K
\end{tablenotes}
\end{threeparttable}
\end{sidewaystable*}

Tables~\ref{table:2} and~\ref{table:3} list the computed equilibrium
lattice parameter, single-crystal elastic constants, and
polycrystalline moduli derived from the single-crystal data of
$\alpha$-Fe along with those from previous
calculations~\cite{liu2005structure,Zhang:2010b,Sha:2006,fellinger2017ab} and
available experimental
data~\cite{Rayne:1961,Adams:2006,Speich:1972}.

The obtained equilibrium lattice constant of bcc Fe is 2.838\AA,
which is consistent with the results from
Refs.~\cite{liu2005structure,Zhang:2010b,Sha:2006,fellinger2017ab}. All the
theoretical results underestimate the experimental value obtained at
ambient conditions (2.866\,\AA)~\cite{Rayne:1961} by approximately
1-2\,\%. The small deviation is partly due to thermal expansion
neglected in theory and also to the employed exchange-correlation
approximation.

The present theoretical method overestimates $C_{11}$ and
underestimates $C_{44}$ with respect to the experimental low-temperature values, whereas the computed $C_{12}$ is only slightly
larger in comparison to the averaged experimental single-crystal
data~\cite{Rayne:1961,Adams:2006}; see Table~\ref{table:2}.
It should be pointed out that the underestimation of the equilibrium
lattice parameter and the overestimation of $C_{11}$ are consistent
with each other, i.e., an expanded lattice parameter is expected to
result in a smaller $C_{11}$ for bcc Fe~\cite{Li:2017}. Considering
the four quoted theoretical lattice parameters and single-crystal
elastic constants, our $C_{11}$ is very close to the average
value (291.2\,GPa), whereas the lattice parameter and $C^\prime$
overestimate the averaged data by 0.011\,\AA{} and 5.3\,GPa,
respectively. Our results for $C_{12}$ and $C_{44}$ are 
smaller than the averages over the other theoretical values, 147.0\,GPa and 107.9\,GPa, respectively,
but $C_{12}$ is closer to the experimental results compared to the
other calculations.
Our computed $C_{44}$ of 106.5\,GPa is similar to the three previous DFT results from Refs.\mbox{~\cite{Zhang:2010b,liu2005structure,fellinger2017ab}} of 98-106.7\,GPa. 
These four theoretical results lie in a narrow interval somewhat below the low-temperature experimental values of 120.8-121.9\,GPa, indicating a systematic deviation from the experimental value, which may be due to the exchange-correlation approximation. The deviation from the experimental value is, however, within the error bar typically found in elastic constant calculations for transition metals\mbox{~\cite{Soederlind:1993}}.

Turning to the polycrystalline moduli of pure Fe, the calculated values for $B$ and $G$ overestimate the isotropic moduli derived from the low-temperature experimental single-crystal data~\cite{Rayne:1961,Adams:2006} and those measured for polycrystalline $\alpha$-Fe~\cite{Speich:1972}; see Table~\ref{table:2}.
The larger bulk modulus is evidently a result of the overestimated $C_{11}$, whereas errors in both $C_{11}$ and $C_{44}$ contribute to the deviation observed for $G$. Young's modulus, Poisson's ratio, and the $B/G$ ratio have errors smaller than or similar to those of $B$ and $G$, depending on error cancellation.

For the Fe$_{91}$Cr$_{9}$ solid solution, the computed lattice parameter and single-crystal elastic constants are listed in Table~\ref{table:2} along with the polycrystalline data in Table~\ref{table:3}.
The addition of 9 at.\% chromium increases the lattice parameter, decreases $C_{12}$ slightly more pronounced than $C_{11}$, thus leading to a larger $C^\prime$, and increases $C_{44}$.
For the derived polycrystalline data, we found that $G$ and $E$ increase with Cr addition, whereas $B$, $\nu$, and $B/G$ shrink.
These trends are in close agreement with a previous theoretical investigation for random Fe$_{90}$Cr$_{10}$ by Zhang \emph{et al.}~\cite{Zhang:2010b} and qualitatively consistent with the experimental findings by Speich \emph{et al.}~\cite{Speich:1972} for 10.72\,at.\% Cr containing Fe-Cr poly crystals.
Namely, from our calculations, we find that the rate of change for $B$ is about $-1.29\,$GPa per atomic percent Cr (determined from Fe$_{91}$Cr$_{9}$ relative to pure Fe) in comparison to $-0.74$\,GPa/at.\%Cr and $-0.83$\,GPa/at.\%Cr at 77\,K and 298\,K, respectively (determined from Ref.~\cite{Speich:1972}, Fe$_{89.28}$Cr$_{10.72}$ relative to pure Fe). Similarly, the computed rates of change for $G$ and $E$ are $+0.81$\,GPa and $+1.50$\,GPa per atomic percent Cr, respectively, which are larger than the experimental data, $+0.28$\,GPa/at.\%Cr and $+0.47$\,GPa/at.\%Cr (at 77\,K), and $+0.35$\,GPa/at.\%Cr and $+0.62$\,GPa/at.\%Cr (at 298\,K), respectively.

Overall, the present theoretical tool captures the compositional effect of 9\,at.\% Cr on the lattice parameter, single-crystal elastic constants, and polycrystalline moduli of bcc Fe reasonably well,  providing support for our systematic study on the effects of other alloying elements on the properties of the Fe$_{91} $Cr$_{9}$ host.

\begin{table*}
\caption{\label{table:3}Theoretical (0\,K) and experimental (temperature as stated in the table notes) polycrystalline elastic moduli, Poisson's ratio, and $B/G$ ratio for pure Fe, Fe$ _{91} $Cr$ _{9}$, CLAM/CLF-1, F82H, and EUROFER97. Abbreviations are as in Table~\ref{table:2}.}
\begin{threeparttable}
\begin{tabular}{*{3}{l}d{3.1}*{3}{l}}
\toprule
Material & Method & $B$ (GPa) & \multicolumn{1}{r}{$G$ (GPa)} & $E$ (GPa) & $\nu$ & $B/G$ \\
\midrule
Fe & This work & 189.4 & 93.7  & 241.3 & 0.288 & 2.02 \\
& EMTO~\cite{Zhang:2010b} & 193.9 & 94.1 & 243.0 & 0.291 & 2.06 \\
& PAW~\cite{liu2005structure}\tnote{a} & 194 & 87 & 226 & 0.306 &  2.24 \\
& PAW~\cite{fellinger2017ab}\tnote{a} & 191 & 83 & 218 & 0.310 & 2.30\\​
& FPLMTO~\cite{Sha:2006}\tnote{a} & 201 & 129 & 318 & 0.236 & 1.56 \\
& Expt.~\cite{Rayne:1961}\tnote{a} & 173.1 & 87.5 & 224.7 & 0.284 & 1.98 \\
& Expt.~\cite{Adams:2006}\tnote{a} & 170.4 & 86.2 & 221.2 &  0.283 & 1.98 \\
& Expt.~\cite{Speich:1972} (77\,K)\tnote{b}  & 172 & 85 &  219 & 0.288 & 2.02\\
& Expt.~\cite{Speich:1972} (298\,K) & 166.0 & 80.7 & 208.2 & 0.291 & 2.06\\
\hline
Fe$_{91}$Cr$_{9}$ & This work & 177.8 & 101.0 & 254.8 & 0.261 & 1.76 \\
Fe$_{90}$Cr$_{10}$ & EMTO~\cite{Zhang:2010b} & 181.2 & 102.3 & 258.3 & 0.262 & 1.77 \\
Fe$_{89.28}$Cr$_{10.72}$ & Expt.~\cite{Speich:1972} (77\,K)\tnote{b}  & 164 & 88 & 224 & 0.272 & 1.87 \\
& Expt.~\cite{Speich:1972} (298\,K) & 157.1 & 84.5 & 214.8 & 0.282 & 1.96 \\
\hline
CLAM/CLF-1& This work & 179.2 & 102.0 & 257.3 & 0.261 & 1.76 \\
& Expt.~\cite{Xu:2014} (298\,K) & 181.7 & 83.8 & 218.0 & 0.300 & 2.13 \\
F82H & This work & 178.5 & 100.5 & 253.9 & 0.263 & 1.78 \\
& Expt.~\cite{Tavassoli:2002} (298\,K) & 171.5 & 84.3 & 217.3 & 0.291 & 2.06 \\
EUROFER97 & This work & 179.0 & 102.0 & 257.1 & 0.261 & 1.76 \\
\bottomrule
\end{tabular}
\begin{tablenotes}
\item[a] Obtained from corresponding single-crystal data in Table~\ref{table:2}
\item[b] $G$ and $E$ taken from figure, $B$ and $\nu$ obtained through Eqs.~\eqref{eq:polyEnu}
\end{tablenotes}
\end{threeparttable}
\end{table*}

\subsection{Lattice parameter and elastic properties of CLAM/CLF-1, F82H, and EUROFER97}

We briefly discuss our theoretical findings for CLAM/CLF-1, F82H, and EUROFER97 and compare them to the available experimental data for the polycrystalline moduli (only CLAM/CLF-1 and F82H); see Tables~\ref{table:2} and~\ref{table:3} for the data. A detailed account of the alloying effects is presented in Secs.~\ref{sec:alloyingeffectlatt} and~\ref{sec:alloyingeffectelas}.

Because of their close chemical compositions (Table~\ref{table:1}), the solid solution description of these three RAFM steels results in nearly identical equilibrium lattice parameters and rather similar elastic properties.
Relative to pure Fe, the addition of the four considered alloying elements Cr, W, V, and Mn increases the lattice constant by approximately 0.5\,\%.
The most significant compositional effect among the single-crystal elastic constants is observed for $C_{44}$ with maximum variation 13.9\,GPa for CLAM/CLF-1. 
The absolute alloying effects on $C_{11}$ and $C_{12}$ are smaller throughout, the decrements being 7.3\,GPa (9.5\,GPa/7.5\,GPa) and 11.7\,GPa (11.6\,GPa/11.9\,GPa) for CLAM/CLF-1 (F82H/EUROFER97) with respect to bulk Fe.
It follows that $C^\prime$ experiences a compositional effect in between those of $C_{11}$ and $C_{12}$.
It is evident from the data reported in Table\mbox{~\ref{table:2}} and the discussion in the previous section that the main alloying effect is due to 9\,at.\% Cr, while the other elements ensure a fine tuning of the elastic parameters. This is the primary reason for discussing the alloying effects relative to Fe$_{91}$Cr$_{9}$ in Sec.\mbox{~\ref{sec:discussion}}. 
We refer the reader to Refs.\mbox{~\cite{Olsson:2006,Zhang:2010b,Korzhavyi:2009}} for a more elaborate discussion of the alloying effect of Cr in pure Fe.

Turning to the polycrystalline parameters, the computed bulk moduli of CLAM/CLF-1, EUROFER97, and F82H are smaller than that of pure Fe, but slightly above the value for Fe$_{91}$Cr$_{9}$.
Both the shear modulus and Young's modulus of the RAFM steels are enhanced with respect to Fe and approximately equal the theoretical moduli of Fe$_{91}$Cr$_{9}$.
The available experimental $G$ and $E$ values for CLAM/CLF-1 and F82H differ by less than 1\,GPa and are close to the room-temperature moduli measured for Fe$_{89.28}$Cr$_{10.72}$ (Ref.~\cite{Speich:1972}). Relative to Fe, both experimental compositional effects agree with the present findings, although the theoretical values are systematically larger.
The situation for the bulk modulus is less clear. The two values for CLAM/CLF-1 and F82H differ from each other by approximately 10\,GPa, thus much more than $G$ and $E$ do. This larger difference in $B$ is, however, reproducible from Eqs.~\eqref{eq:polyEnu}, and the same applies to $\nu$. On the other hand, the bulk moduli of CLAM/CLF-1 and F82H are significantly larger than the $B$ reported for the Fe$_{89.28}$Cr$_{10.72}$ binary~\cite{Speich:1972}.

\begin{figure}[tbh]
\resizebox{\columnwidth}{!}{\includegraphics[clip]{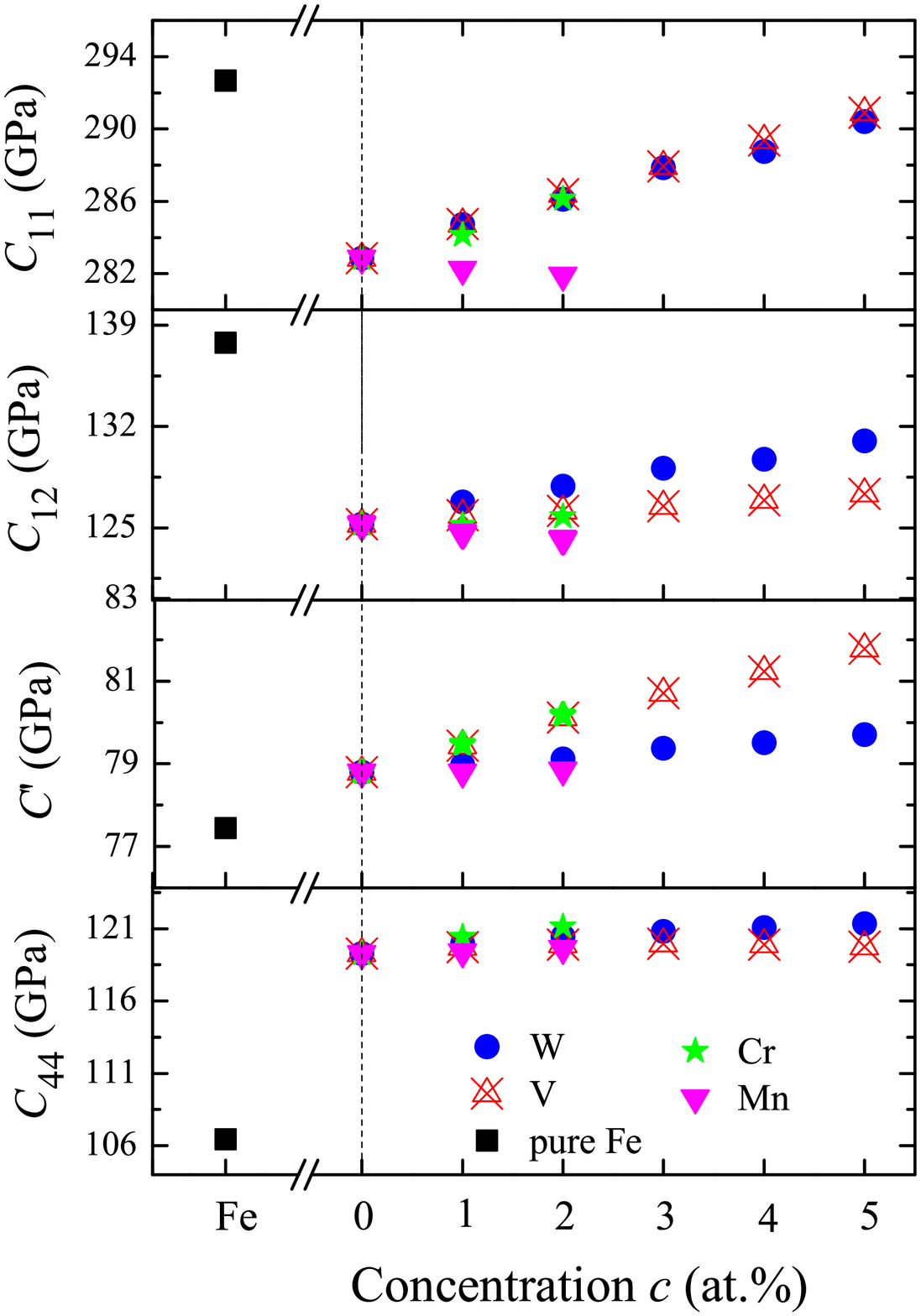}}
\caption{\label{fig:1}Theoretical single-crystal elastic constants $C_{11}$, $C_{12}$, and $C_{44}$, and shear elastic constant $C^\prime$ for bcc Fe and Fe$_{91-c}$Cr$_{9}M_{c}$ alloys, where $c=\bar{x},y,z,\text{ or }u$ for $M$=Cr, W, V, or Mn, respectively. Note that $c=0$ corresponds to the Fe$_{91}$Cr$_{9}$ binary.}
\end{figure}

\subsection{Surface energy and unstable stacking fault energy}

\begin{table*}
    \caption{\label{table:5}Theoretical surface energy $\gamma_{\text{s}(110)}$, USF energy $\gamma_{\text{u}(110)}$, and the ratio $\gamma_{\text{s}(110)}/ \gamma_{\text{u}(110)}$ for pure Fe,  Fe$ _{91} $Cr$ _{9}$, CLAM/CLF-1, F82H, EUROFER97. Available estimated surface energies for Fe is also listed.}
    \begin{threeparttable}
        \begin{tabular}{*{5}{l}}
            \toprule
            Material & Method  &  $\gamma_{\text{s}(110)}$ (J/m$^2$) & $\gamma_{\text{u}(110)}$ (J/m$^2$) & $\gamma_{\text{s}(110)}/ \gamma_{\text{u}(110)}$ \\
            \midrule
            Fe & This work &  2.437 & 1.075 & 2.266 \\
            & Other theory & 2.37~\cite{Blonski:2007}, 2.45~\cite{Punkkinen:2011b}, 2.47~\cite{Wang:2015}  & 0.98~\cite{Mori:2009}, 1.08~\cite{Wang:2015} & 2.287~\cite{Wang:2015} \\
            & Estimate\tnote{a}  &  2.41~\cite{Tyson:1977}, 2.48~\cite{deBoer:1988} & - & - \\
            \hline
            Fe$_{91}$Cr$_{9}$ & This work & 2.443 & 1.108 & 2.204 \\
            \hline
            CLAM/CLF-1 & This work & 2.453 & 1.120 & 2.190 \\
            F82H & This work & 2.445 & 1.112 & 2.198 \\
            EUROFER97 & This work & 2.452 & 1.117 & 2.194 \\
            \bottomrule
        \end{tabular}
        \begin{tablenotes}
            \item[a] semi-empirical 0\,K estimate for average surface facet derived from experimental surface tension (liquid phase) neglecting magnetic entropy in extrapolation
        \end{tablenotes}
    \end{threeparttable}
\end{table*}

We first establish both planar fault parameters for Fe and then discuss our theoretical findings for Fe$_{91}$Cr$_{9}$, CLAM/CLF-1, F82H, and EUROFER97.

The surface energy $\gamma_{\text{s}(110)}$ and the USF energy $\gamma_{\text{u}(110)}$ of solid Fe presented in Table~\ref{table:5} are in close agreement with the other tabulated theoretical results.
As for experimental data, semi-empirical estimations of the surface energy for an average surface facet have been derived from surface-tension data in the liquid phase~\cite{Tyson:1977,deBoer:1988}, whereas USF energies are not accessible to measurements.
Depending on the details of the extrapolation schemes, the 0\,K estimates for $\gamma$ amount to 2.41\,J/m$^2$~\cite{Tyson:1977} and 2.48\,J/m$^2$~\cite{deBoer:1988}. Our calculated value 2.437\,J/m$^2$ agrees well with these estimates of 2.41-2.48\,J/m$^2$, and is similar to the other theoretical results  2.37-2.48\,J/m$^2$\mbox{~\cite{Blonski:2007,Punkkinen:2011b,Wang:2015}}.
It should be noted that magnetic entropy was neglected in both extrapolations~\cite{Tyson:1977,deBoer:1988}, but estimated to increase the $0$\,K estimates by 0.14\,J/m$^2$~\cite{Schoenecker:2015b}.

The calculated planar energies of the CLAM/CLF-1, F82H, and EUROFER97 grades are listed in Table~\ref{table:5}. Their close chemical compositions lead to rather similar surface and USF energies.
It is found that the collective addition of Cr, W, V, and Mn increases the surface energy. Similarly, the USF energy increases relative to Fe. For instance, $\gamma_{\text{u}(110)}$ increases from 1.075 to 1.120\,J/m$^2$ (4.2\,\%) for CLAM/CLF-1, which outweighs the corresponding increase of $\gamma_{\text{s}(110)}$ by 0.7\,\%.
The $\gamma_{\text{s}(110)}$ and $\gamma_{\text{u}(110)}$ of Fe$_{91}$Cr$_{9}$ amount to 2.443\,J/m$^2$ and 1.108\,J/m$^2$, respectively, which are close to the results for the above three RAFM steels. Thus, Cr constitutes a major part of the increased surface and USF energies of CLAM/CLF-1, F82H, and EUROFER97 (Table~\ref{table:5}).

\section{\label{sec:discussion}Discussion}

\subsection{\label{sec:alloyingeffectlatt}Detailed alloying effects on the lattice parameter of Fe$_{91}$Cr$_{9}$}

To learn about the effects of the individual solute atoms, we go beyond their experimental concentrations  and discuss the elastic properties of the ternary random solid solutions Fe$_{91-c}$Cr$_{9}M_c$, where, for brevity, the concentration variable $c$ denotes $c=\bar{x},y,z,\text{ or }u$ for solute $M$=Cr, W, V, or Mn, respectively. Here, $\bar{x}\equiv x - 9$ was introduced, i.e., we use the 9\,at.\% Cr alloy as reference.
The considered concentration ranges are $0 \le y,z \le 5$ at.\,\% and $0 \le \bar{x},u \le 2$ at.\,\% in intervals of 1 at.\,\%. Thus, alloying with W, V, or Mn maintains the Cr concentration at 9\,at.\%, whereas alloying with Cr increases  it  beyond this amount.

The reason for choosing Fe$_{91}$Cr$_{9}$ as the base alloy  is that the alloying effects of Cr, W, V, and Mn in the Fe-Cr host are expected to differ from those in pure Fe, i.e., the alloying effects are not transferable.
This is plausible since nonmonotonic variations of $C_{11}$ and $B$ with Cr addition below 10\,at.\% to bcc Fe were reported previously in Refs.~\cite{Zhang:2010b,Korzhavyi:2009} (and experimental references therein), and an anomalous sign change of $dB/dx$ at approximately 7\,at.\% Cr was ascribed to change in the Fermi surface topology~\cite{Korzhavyi:2009}.

\begin{table*}[tbh]
\caption{\label{table:4}The total rates of change (in boldface) of $a$, $C^{\prime}$, $C_{44}$, and $B$ upon alloying the bcc Fe$_{91}$Cr$_{9}$ solid solution with $M=$ Cr,  W, V, or Mn, and the decomposition into electronic (ele.) and volumetric (vol.) effects according to Eq.~\eqref{eq:roc_ae}.}
\begin{tabular}{ld{3.1}*{9}{r}}
\toprule
 & \multicolumn{1}{c}{$\Delta a/ \Delta c$} & \multicolumn{3}{c}{$\Delta C^\prime/ \Delta c$} & \multicolumn{3}{c}{$\Delta C_{44}/ \Delta c$} & \multicolumn{3}{c}{$\Delta B/ \Delta c$} \\
 \cmidrule(lr){3-5}
 \cmidrule(lr){6-8}
 \cmidrule(lr){9-11}
 & & $\textbf{total}$ & ele. & vol. & $\textbf{total}$ & ele. & vol. & $\textbf{total}$ & ele. & vol. \\
 \cmidrule(lr){3-11}
 $M$ & \multicolumn{1}{c}{($10^{-3}$\AA/at.\%)} & \multicolumn{9}{c}{(GPa/at.\%)}  \\ 
\midrule
Cr & -0.367 & $\textbf{0.692}$ & 0.598 & 0.095   & $\textbf{0.928}$  & 0.742 & 0.186 & $\textbf{0.720}$ & 0.353 & 0.368 \\
W & 5.606 & $\textbf{0.156}$ & 1.600 & -1.444  & $\textbf{0.560}$ & 3.405 & -2.845 & $\textbf{1.415}$ & 7.034 & -5.619 \\
V & 0.536 & $\textbf{0.658}$  & 0.796 & -0.138  & $\textbf{0.311}$ & 0.583  & -0.272 & $\textbf{0.899}$ & 1.436 & -0.537 \\
Mn & -0.587 & $\textbf{0.026}$  & -0.125  & 0.151 & $\textbf{0.182}$ & -0.116 & 0.298 & $\textbf{-0.504}$ & -1.092 & 0.589\\
\bottomrule
\end{tabular}
\end{table*}

\begin{figure}[tbh]
\resizebox{\columnwidth}{!}{\includegraphics[clip]{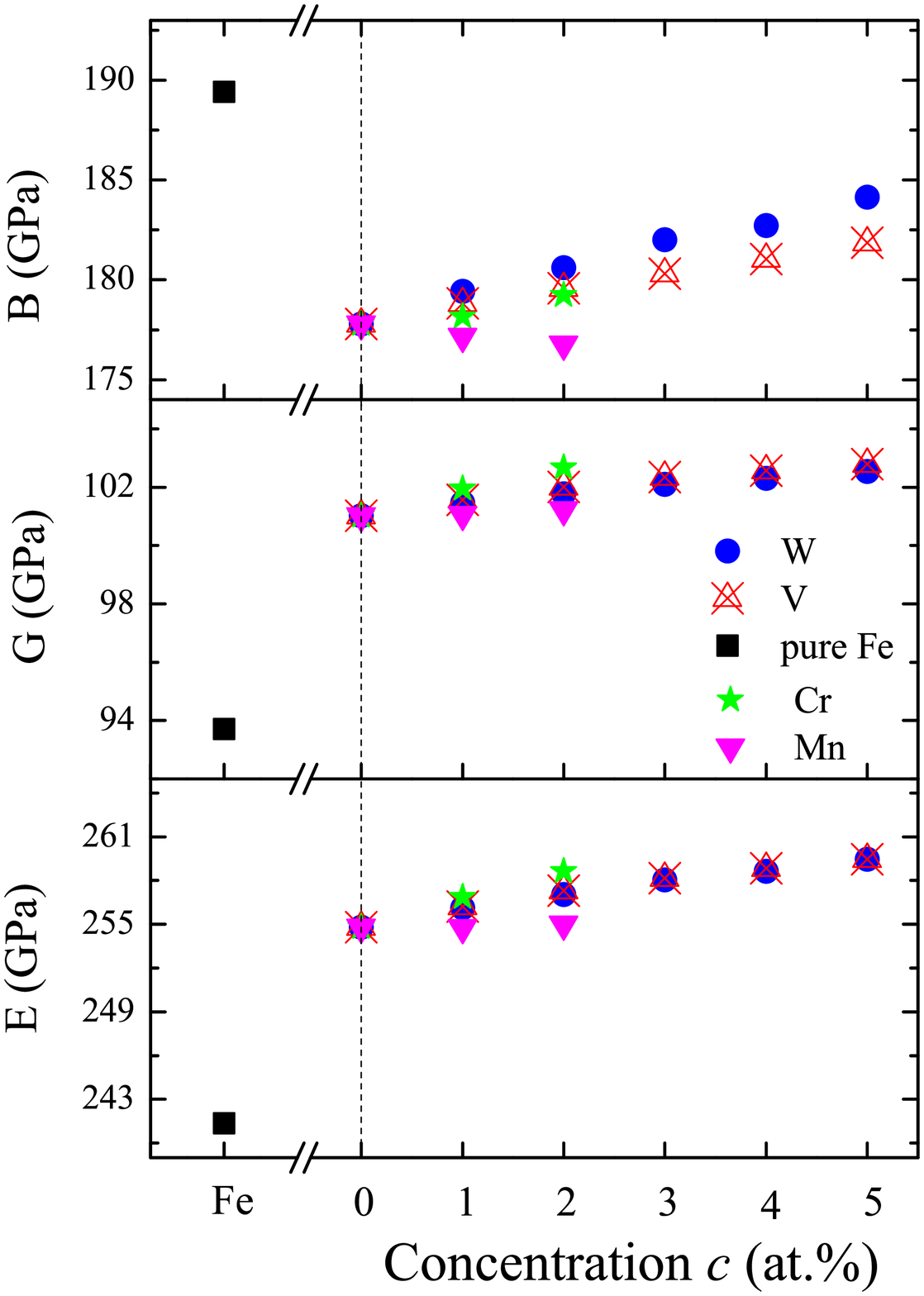}}
\caption{\label{fig:2}Theoretical bulk modulus $B$, shear modulus $G$, and Young's modulus $E$ for bcc Fe and Fe$_{91-c}$Cr$_{9}M_{c}$ alloys, where $c=\bar{x},y,z,\text{ or }u$ for $M$=Cr, W, V, or Mn, respectively. Note that $c=0$ corresponds to the Fe$_{91}$Cr$_{9}$ binary.}
\end{figure}

Our results for Fe$_{91-c}$Cr$_{9}M_c$ are shown in Figs.~\ref{fig:1} and~\ref{fig:2}. In these figures, the data for pure Fe from Tables~\ref{table:2} and~\ref{table:3} are also included to illustrate the alloying effect of the solute $M$ on the properties of the Fe$_{91}$Cr$_{9}$ host relative to the changes induced by solving 9 at.\,\% Cr in Fe.
Overall speaking, W (Mn) has on average the largest (smallest) alloying coefficients among the four considered solutes.
For both Cr and Mn negative changes in the lattice parameter arise, whereas V and, in particular, W cause an expansion of the volume. 

\begin{figure}[tbh]
\resizebox{\columnwidth}{!}{\includegraphics[clip]{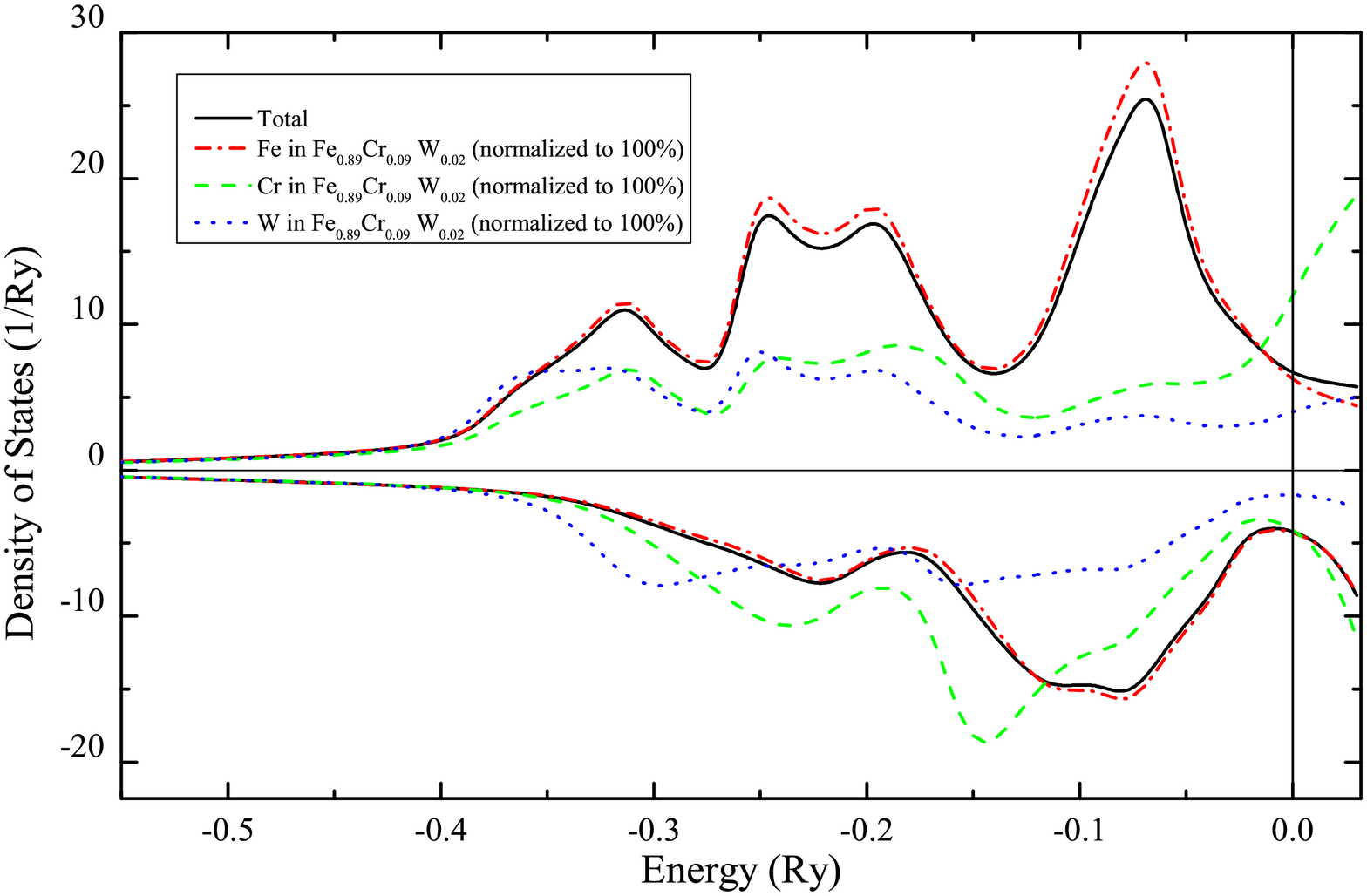}}
\caption{\label{fig:dos}Total DOS for bcc Fe$_{89}$Cr$_{9}$W$_2$ alloy and partial contributions of Fe, Cr, and W. The contributions of the elements were normalized to 100 at.\,\% in order to ease the comparison. Majority and minority states carry a positive and negative sign, respectively. The energy is plotted relative to the Fermi energy.}
\end{figure}

In order to understand the electronic origin of the large volume-increasing effect of W, we focus on the electronic structure of Fe$_{89}$Cr$_{9}$W$_2$ shown in Fig.\mbox{~\ref{fig:dos}}.
As in previous work on metallic bonding, we interpret the electronic structure of the alloy in terms of 'common-band' and 'split-band' behavior\mbox{~\cite{Staunton:1998,Pettifor:1995}}. Roughly speaking, in the 'common-band' model the energy separation $\Delta \epsilon$ of the $d$ band centers associated with each type of atom relative to the $d$ bandwidth $w$ is small ($\Delta \epsilon/w \ll 1$), while in 'split-band' behavior this energy separation is not small. It should be noted that the actual alloy electronic band structure is a (complicated) mixture of both models.
Bulk elemental Fe and Cr have similar $d$ bandwidths of $\approx 6$\,eV and both are exchange-split. The band centers of the minority spin states of Fe and the majority spin states of Cr are close in energy relative to their common bandwidth and align in the Fe-rich alloy to form a 'common-band'-like minority spin band. 
Due to the different exchange splittings of Fe and Cr, the majority spin electron DOS of the alloy shows pronounced 'split-band' behavior. This is evident since the $d$ states of Cr form a Lorentzian-type virtual bound state (VBS) above the majority $d$ states of Fe  and arises from the hybridization of Cr $d$ states with the $s$ and $p$ states of Fe\mbox{~\cite{Dederichs:1991}} (the tail of the VBS also appears in Fe$_{89}$Cr$_{9}$W$_2$ above the Fermi level; see Fig.\mbox{~\ref{fig:dos}}). 
As a result, the Cr moment is aligned oppositely to the Fe moment.
Bulk W has a $d$ bandwidth of $\approx 12$\,eV and there is no exchange splitting. Due to the different bandwidths of W and the Fe$_{91}$Cr$_{9}$ host, the electronic structure of the ternary Fe$_{89}$Cr$_{9}$W$_2$ alloy is characterized by strong 'split-band' formation in both spin channels, i.e., W and the Fe$_{91}$Cr$_{9}$ matrix tend to form states that reside mostly on W or Fe$_{91}$Cr$_{9}$. This is, for instance, evident from the spectral weight near the bottom of the alloy DOS mainly associated with the W sites. 
Hybridization effects are stronger for the minority $d$ alloy states owing to the smaller energy separation of the $d$ band centers of W and Fe$_{91}$Cr$_{9}$. Fe induces an exchange splitting on W sites in order to reduce the kinetic energy (the moments of W are aligned oppositely to the Fe moments). 
Similar to Cr, the $d$ states of W also form 'split-band' VBSs situated above the majority $d$ states of Fe\mbox{~\cite{Dederichs:1991}} (located above the maximum energy plotted in Fig.\mbox{~\ref{fig:dos}}).

We find that the $d-d$ hybridization between the states associated with W atoms and the Fe-rich matrix leads to a charge transfer between the $d$ states at W and Fe sites, which reduces the valence electron number around a tungsten site from approximately 6 in bulk W to 4.5 in the alloy. 
As a result of the drop in the $d$-electron number, the attractive partial $d$-pressure\mbox{~\cite{Pettifor:1987b,Pettifor:1995}} from the W sites decreases in absolute value to the level between that of Hf and Ta, which have an averaged valence electron number of approximately 4.5 and an averaged bcc lattice parameter of $(3.534\textrm{\,\AA}+3.327\textrm{\,\AA})/2 = 3.431$\,\AA{} (as determined by EMTO and PBE).
Vegard's rule then gives for the variation of the lattice parameter upon addition of W with 4.5 valence electrons to Fe $(3.431-2.838)\textrm{\,\AA}/100$\,at.\,\%\,W$\,=\,$0.0059\,\AA/at.\,\%\,W. This slope agrees well with the calculated variation upon addition of 2\,at.\% W to Fe$_{91}$Cr$_{9}$, i.e., $(2.860-2.849)\textrm{\,\AA}/2$\,at.\,\%\,W $=$0.0055\,\AA/at.\%\,W.
We notice that Vegard's rule for the variation of the lattice parameter using the equilibrium lattice parameter of pure bcc W (i.e. omitting the hybridization effects) is $(3.191-2.838)\textrm{\,\AA}/100$\,at.\%\,W $\,=\,$0.0035\,\AA/at.\%\,W, which is substantially smaller than the previous figure obtained by taking into account the interaction between W and the Fe-rich matrix.

\subsection{\label{sec:alloyingeffectelas}Detailed alloying effects on elastic properties of Fe$_{91}$Cr$_{9}$}

The elastic constants $C_{11}$ and $C_{12}$ increase with W, Cr, or V addition, but slightly decrease with Mn (see \mbox{Fig.~\ref{fig:1}}). For instance, for 2 at.\,\% Cr, W, V, or Mn addition, the variations of $C_{11}$ are approximately 1.2\,\%, 1.2\,\%, 1.3\,\%, or -0.3\,\%, respectively.
The resulting trends for $C^\prime$ are all positive, i.e., a variation of approximately 1.8\,\%, 0.4\,\%, 1.7\,\%, or 0.1\,\%  with increasing the amount of Cr, W, V, or Mn, respectively, indicating that the compositional trends on $C_{11}$ outweigh the ones on $C_{12}$.
Clearly, $C_{44}$ increases with $c$ for all solutes and the largest slope is observed for Cr. 
The predicted influences on the bulk modulus are a strong increase for W, weaker increases for Cr and V, and a negative slope for Mn, respectively.
The alloying effects on $G$ are systematically small, $E$ increases with W, V, or Cr addition, whereas the impact of Mn on $E$ is negligible.
The present alloying effects of W on Fe$_{91}$Cr$_{9} $ are consistent with the previous findings for Fe-Cr-W alloys~\cite{Xu:2014}.

The dilute alloying effects of the individual solutes on the lattice parameter and selected elastic properties of Fe$_{91}$Cr$_{9}$ are quantified in Table~\ref{table:4} by their rate of change $\Delta \mathcal{C} (c)/ \Delta c$, where $\mathcal{C}$ stands for $a$, $C^\prime$, $C_{44}$, or $B$. These parameters were evaluated by averaging the compositional effect in the concentration interval $0\le c \le 2$ at.\,\%.
A more detailed analysis of the alloying trends can be achieved by decomposing the predicted total effects into separate effects, such as a volumetric contribution due to the lattice expansion/contraction accompanying solute addition and a purely electronic contribution due to alloying, \emph{viz.}
\begin{align}
\underbrace{\frac{\Delta \mathcal{C} }{\Delta c} }_{\text{total}} &= \underbrace{ \left(\frac{\Delta \mathcal{C}  }{\Delta c}\right)_{a} }_{{\text{electronic}}} + \underbrace{ \left(\frac{\Delta \mathcal{C} }{\Delta a}\right)_{c}\frac{\Delta a}{\Delta c} }_{\text{volumetric}}.
\label{eq:roc_ae}
\end{align}
Apart from the already determined total rates of change and the change in the equilibrium lattice parameter accompanying alloying $\Delta a /\Delta c$, we also evaluated the volumetric effect at constant concentration for the Fe$_{91}$Cr$_{9}$ host. The coefficients are $(\Delta C_{44}/\Delta a)_{c}=-507.450$\,GPa/\AA{}, $(\Delta C'/\Delta a)_{c}=-257.495$\,GPa/\AA{}, $(\Delta B/\Delta a)_{c}=-1002.342$\,GPa/\AA{}, indicating that the volumetric contribution leads to a softening of these elastic parameters as the lattice expands.
The electronic effects were evaluated as to balance the total and volumetric effects in Eq.~\eqref{eq:roc_ae}.

It should be remarked that an analysis similar to Eq.~\eqref{eq:roc_ae} was used before~\cite{Speich:1972,Ghosh:2002,Li:2017,fellinger2017ab}, 
but each approach differs in terms of methodological details. 
On the one hand, the electronic contribution was evaluated as the difference between the total and volumetric terms in Refs.\mbox{~\cite{Speich:1972,Ghosh:2002,Li:2017}} and this work, whereas it was computed separately in Ref.\mbox{~\cite{fellinger2017ab}}. Fellinger \emph{et al.} showed\mbox{~\cite{fellinger2017ab}} that the sum of the separately computed volumetric and electronic terms agrees with the directly computed total rate of change. 
On the other hand, the volumetric term was either determined directly in Ref.\mbox{~\cite{Li:2017}} and this work, or indirectly with the help of the pressure derivative of $\mathcal{C}$\mbox{~\cite{Speich:1972,Ghosh:2002}} or through the solute induced stress in the host matrix\mbox{~\cite{fellinger2017ab}}.

The electronic and volumetric contributions to the rates of change in $C'$, $C_{44}$, and $B$ upon alloying the Fe$_{91}$Cr$_{9}$ solid solution with Cr, W, V, or Mn are also listed in Table~\ref{table:4}. Evidently, all the electronic and volumetric contributions are found to have opposite signs with the exception of Cr, and the signs of all volumetric effects on the slopes of $C'$, $C_{44}$, and $B$ are opposite to the sign of $\Delta a/ \Delta c$.
In addition, all the volumetric rates of change involving W are approximately a factor of ten larger than those of the other solid solution elements. Although the volumetric rates of change are large in W, they are mostly compensated by the electronic effects and they nearly balance each other in the case of $\Delta C^\prime / \Delta c$.
For Cr and V, the electronic effect on the slope of $C'$ is approximately six times larger in magnitude than the volumetric part, whereas both effects have similar absolute values for W and Mn. This grouping does, however, not universally occur for all three considered elastic parameters.

It should be noted that a moderate positive correlation between $\Delta a/ \Delta c$ and the slope of $C_{12}$ (Fig.~\ref{fig:1}) for Cr, W, V, or Mn addition exists. Similar, but progressively weaker positive correlations are also observed between $\Delta a/ \Delta c$ and $\Delta B/ \Delta c$, and between $\Delta a/ \Delta c$ and $\Delta C_{11}/ \Delta c$.
$\Delta C_{44}/\Delta c$ does not exhibit such obvious ordering.
It should further be noted that there is no obvious correlation between the value of $C_{44}$ of the solute in its ground state and its effect on $C_{44}$ of Fe$_{91}$Cr$_{9}$. For instance, the values of $C_{44}$ for V and Cr are lower than that of Fe$_{91}$Cr$_{9}$ from theory and experiment~\cite{Gao:2013}, but both solutes exhibit a positive $\Delta C_{44}/\Delta c$.

\begin{table}[tbh]
    \caption{\label{table:relativechange}The theoretical relative alloying effects (in \%) for CLAM/CLF-1, F82H, EUROFER97, and a test alloy with composition Fe$_{84.4}$Cr$_{9.6}$W$_{2}$V$_{2}$Mn$_{2}$ with respect to Fe$_{91}$Cr$_{9}$.
    The numbers listed first were obtained by means of Eq.~\eqref{eq:superpos} ($C'$, $C_{44}$, and $B$) and through the Hill average and Eqs.\mbox{~\eqref{eq:polyEnu}} ($G$, $E$, and $\nu$).
    The reference data in parentheses were directly derived from the computed data (Tables~\ref{table:3} and~\ref{table:5}). }
    \begin{tabular}{l*{5}{d{3.2}}d{4.3}}
        \toprule
        Grade & \multicolumn{1}{r}{$\Delta C'/ C'$}  & \multicolumn{1}{r}{$\Delta C_{44}/C_{44}$}  & \multicolumn{1}{r}{$\Delta B/B$}  & \multicolumn{1}{r}{$\Delta G/G$} & \multicolumn{1}{r}{$\Delta E/E$} & \multicolumn{1}{r}{$\Delta \nu/\nu$}\\
        \midrule
        CLAM/CLF-1 & 1.09  & 0.91  & 0.75  & 0.98  & 0.94  & -0.175 \\
        & (1.09) & (0.95) & (0.78) & (1.01) & (0.97) & (-0.175) \\ 
        F82H &  -0.41  & -0.34  & 0.22  & -0.37  & -0.28  & 0.456  \\
        & (-0.36) & (-0.58) & (0.40) & (-0.49) & (-0.35) & (0.687) \\
        EUROFER97 & 1.05  & 0.78  & 0.67  & 0.89  & 0.85  & -0.166  \\
        & (1.08) & (0.84) & (0.66) & (0.94) & (0.89) & (-0.212) \\
        Test alloy & 2.62  & 2.25  & 2.29  & 2.40  & 2.38  & -0.082   \\
        & (2.16) & (1.34) & (2.10) & (1.67) & (1.74) &(0.330) \\
        \bottomrule
    \end{tabular}
\end{table}

Having established the electronic and volumetric contributions to the total rates of change, we now shed some light on the origin of the strong increase of the bulk modulus upon adding W to the Fe$_{91}$Cr$_{9}$ matrix as discussed at the beginning of this subsection.
It should be noted that the volumetric contribution to the rate of change of the bulk modulus is directly related to the lattice parameter change $\Delta a /\Delta c$. Adopting for simplicity the Murnaghan equation of state~\mbox{\cite{murnaghan1944compressibility}}, this relationship can be expressed as
\begin{align}
 \left(\frac{\Delta B }{\Delta a}\right)_{\text{volumetric}}  &=  \frac{-3BB^\prime}{a} \frac{\Delta a}{\Delta c}. \label{eq:Bvol}
\end{align}
Here, $B^\prime$ is the pressure derivative of the bulk modulus. The coefficient $-3BB^\prime/a$ for Fe$_{91}$Cr$_{9}$ amounts to -837\,GPa/\AA{} and is similar to the one obtained by approximating $\partial B /\partial a $ through finite differences, $(\Delta B/\Delta a)_{c}=-1002.342$\,GPa/\AA{}. In comparison, $-3BB^\prime/a = -1124$\,GPa/\AA{} for pure bcc Fe.
Thus, the large, negative volumetric contribution of W to the rate of change of the bulk modulus emerges from its strong volume-increasing effect. The volume-increasing effect was in turn explained previously using electronic structure arguments in connection with Vegard's rule.
We note in passing that the correlation Eq.\mbox{~\eqref{eq:Bvol}} applies to the other solutes as well.

Turning to the electronic effect, we recall that the valence electron number of a tungsten site in Fe$_{89}$Cr$_{9}$W$_{2}$ is about 4.5 due to the strong hybridization as discussed above. We proceed as before and estimate the bulk modulus of W with 4.5 valence electrons at the Fe$_{91}$Cr$_{9}$ equilibrium volume from the arithmetically averaged $a$, $B$, and $B^\prime$ data for Hf and Ta calculated at their bcc equilibrium volumes through EMTO and PBE. Using again the Murnaghan equation of state, we arrive at a bulk modulus of $810.4$\,GPa. 
Employing the rule of mixture (analogous to Vegard's rule in the case of lattice parameter), this bulk modulus corresponds to an electronic effect $(\Delta B / \Delta c)_a$ of $(810.4-177.8)$\,GPa/100\,at.\% W$\,=\,$6.3\,GPa/at.\% W.  This value is consistent with the value of 7.034\,GPa/at.\% W derived from Eq.\mbox{~\eqref{eq:roc_ae}}.
Thus, the electronic contribution to the rate of change of the bulk modulus upon addition of W to Fe$_{91}$Cr$_{9}$ can be captured by the rule of mixture considering the right end members, namely that the solute has 4.5 valence electrons and is compressed to the equilibrium volume of Fe$_{91}$Cr$_{9}$. We notice again that a straightforward application of the rule of mixture employing the ground state bulk modulus of W cannot explain the observed trend.

Knowing the element specific rates of change $\Delta \mathcal{C} / \Delta c$ in Fe$_{91}$Cr$_{9}$, we propose the following linear superposition, which may be used to predict the properties of the multicomponent solid solution Fe$_{91-\bar{x}-y-z-u}$Cr$_{9+\bar{x}}$W$_{y}$V$_{z}$Mn$_{u}$, for convenience referred to as RAFM,
\begin{equation}
\mathcal{C}_{\text{RAFM}}= \mathcal{C}_{\text{Fe}_{91}\text{Cr}_{9}}+ \sum_{c=\bar{x},y,z,u} c \frac{\Delta \mathcal{C}}{\Delta c}.
\label{eq:superpos}
\end{equation}
The equation is particularly useful since the specific chemical concentrations of Cr, W, V, and Mn in the three RAFM steels listed in Table\mbox{~\ref{table:1}} are not sharp\mbox{~\cite{Huang:2013}}.

In order to establish the range of validity of this linear description,
we put the above equation to the test by evaluating $C^\prime$, $C_{44}$, and $B$ for the CLAM/CLF-1, F82H, and EUROFER97 compositions according to Table~\ref{table:1}. We arrive at $C^\prime = 79.7$, 78.5, and $79.6$\,GPa, $C_{44} = 120.4$, 118.9, and 120.2\,GPa, and $B= 179.1$, 178.2, and 179.0\,GPa for  CLAM/CLF-1, F82H, and EUROFER97, respectively.
Using the theoretical elastic parameters of Fe$_{91}$Cr$_{9}$ as reference, we compare in Table~\ref{table:relativechange} the relative alloying effects evaluated through Eq.~\eqref{eq:superpos} with those derived from the direct calculations (Tables~\ref{table:3} and~\ref{table:5}).
The agreement is generally good; the largest error amounts to 0.3\,GPa (0.3\,\% deviation in the alloying effect) for $C_{44}$ of F82H.
The relative alloying effects for $C^\prime$ and $C_{44}$ are negative in the case of the F82H grade due to Cr deficiency with respect to Fe$_{91}$Cr$_{9}$ and large specific rates of change. This trend is not overridden by adding W, V, and Mn when obtaining the nominal composition of F82H. The relative alloying effect of $B$ for F82H is, however, positive, which follows mainly from  the large $\Delta B/\Delta c$ of W. In contrast, CLAM/CLF-1 and EUROFER97 contain more than 9 at.\,\% Cr and have positive alloying effects throughout (Table\mbox{~\ref{table:relativechange}}).

We also list the relative alloying effects of $G$, $E$, and $\nu$ in Table\mbox{~\ref{table:relativechange}}, which were obtained through the Hill average 
and Eqs.\mbox{~\eqref{eq:polyEnu}} from the single-crystal rates of change. The validation against the directly calculated data tabulated in Table\mbox{~\ref{table:relativechange}} for the CLAM/CLF-1, F82H, and EUROFER97 compositions shows that the accuracy of estimating $G$, $E$, and $\nu$ is comparable to the one of $C^\prime$, $C_{44}$, and $B$.

Based on the data validation presented in the previous two paragraphs, we conclude that the linear description from Eq.\mbox{~\eqref{eq:superpos}} is adequate if $\sum_{|\bar{x}|,y,z,u}\le 2.4$ at.\,\%. This limit corresponds to the summed up maximum concentrations of Cr, W, V, and Mn (Cr content relative to 9 at.\,\%) contained in any of the three considered RAFM steels (Table\mbox{~\ref{table:1}}).

Finally, it is instructive to briefly discuss an additional estimate of the elastic parameters for a bcc, quinary RAFM alloy, which composition falls outside this range of validity.  
Without loss of generality, we chose the composition Fe$_{84.4}$Cr$_{9.6}$W$_{2}$V$_{2}$Mn$_{2}$ ($\sum_{|\bar{x}|,y,z,u} = 6.6$ at.\,\%) and the computed relative alloying effects are included in Table\mbox{~\ref{table:relativechange}}.
Equation~\eqref{eq:superpos} yields 80.9, 122.0, and 181.9\,GPa for $C^\prime$, $C_{44}$, and $B$, respectively, whereas the directly calculated values are 80.5, 120.9, and 181.5\,GPa. 
The relative alloying effect of $C_{44}$ carries the largest error, i.e., an increase of 2.62\,\% is predicted from Eq.\mbox{~\eqref{eq:superpos}} in comparison to a 1.34\,\% increase from the direct computation.
This error propagates to the polycrystalline moduli derived from it and is mainly responsible for the accuracy of predicting $G$. Ultimately, Eq.\mbox{~\eqref{eq:superpos}} does not predict the correct sign of the alloying effect on the Poisson ratio for Fe$_{84.4}$Cr$_{9.6}$W$_{2}$V$_{2}$Mn$_{2}$. The underlying reason is the linear description, which insufficiently approximates nonlinear alloying effects of the individual solutes that occur when their concentration increases, as well as those arising from the interaction of the solutes.

\begin{figure}[tbh]
\resizebox{\columnwidth}{!}{\includegraphics[clip]{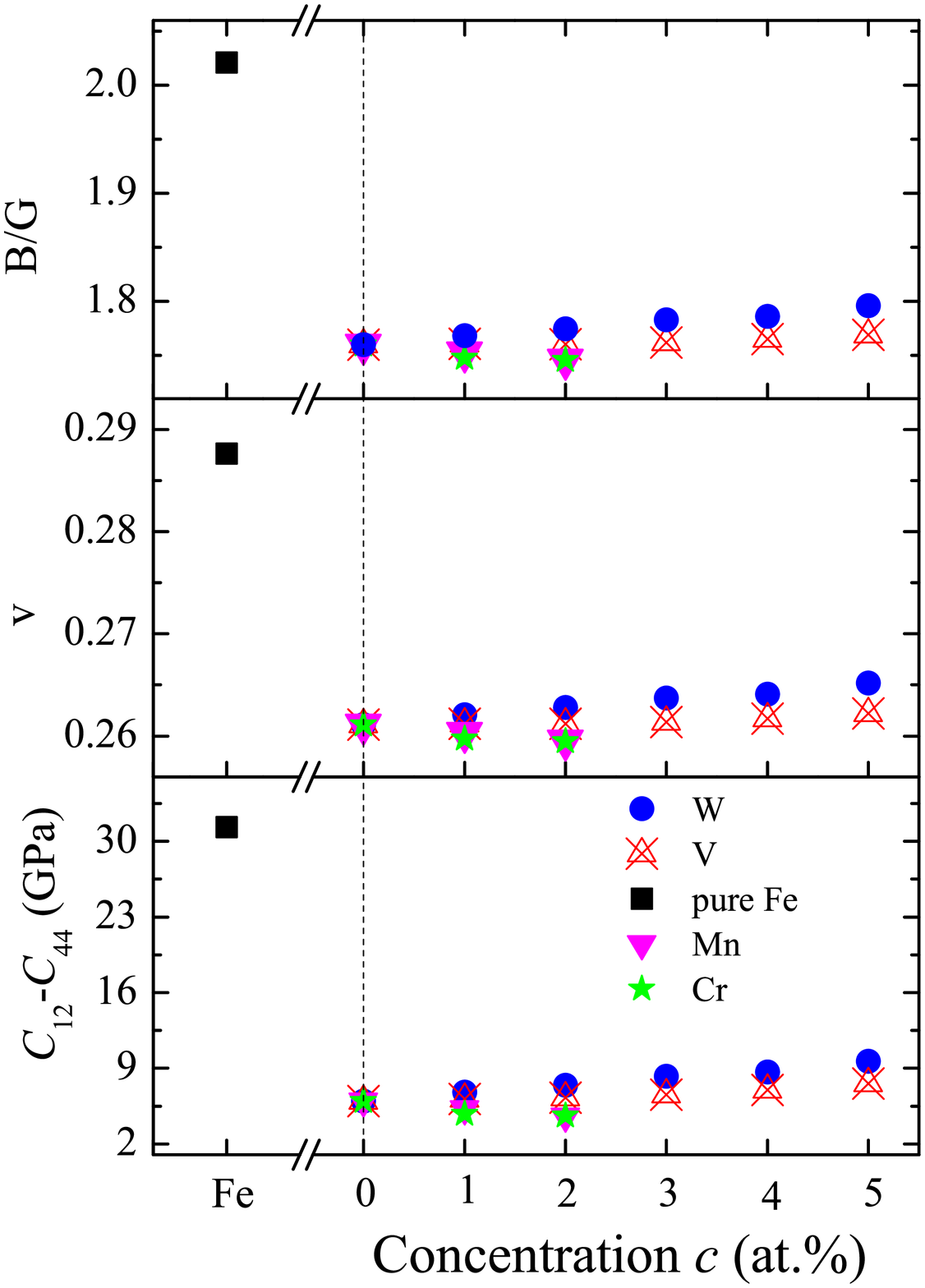}}
\caption{\label{fig:3}Theoretical Pugh ratio $B/G$, Poisson's ratio $\nu$, and Cauchy pressure $C_{12}-C_{44}$ for bcc Fe and Fe$_{91-c}$Cr$_{9}M_{c}$ alloys, where $M$=Cr, W, V, or Mn and $c=\bar{x},y,z,\text{ or }u$. Note that $c=0$ corresponds to the Fe$_{91}$Cr$_{9}$ binary.}
\end{figure}

\subsection{\label{sec:ductility}Ductility}

Ductility is crucial for the performance of a structural material, 
and extrinsic and intrinsic aspects of a material's ductility may be addressed during alloy design. 
Extrinsic features relate to the microstructure of the alloy (e.g., grain size and precipitate distribution). 
Here, we focus on the intrinsic aspect related to the properties of the perfect crystal.

The assessment of intrinsic ductility for crystalline metallic materials may be separated into two classes of methods: empirical relationships and phenomenological theories. Common to both approaches is the attempt to capture the
competition between the ease to brittle cleavage fracture versus the ease to plastic flow.
Two widely used empirical indicators are Pugh's criterion~\cite{Pugh:1954} and the Cauchy pressure~\cite{Pettifor:1992}. Pugh's criterion employs the ratio of polycrystalline bulk modulus to shear modulus $B/G$ and is expected to hold below temperatures of one third of the melting point,  
Following Pugh, ductile (brittle) materials typically have a large (small) $B/G$.
The Cauchy pressure defined as $C_{12}-C_{44}$ was suggested to reflect the nature of the atomic bonding. According to Pettifor\mbox{~\cite{Pettifor:1992}}, a negative Cauchy pressure indicates a more angular character of the bonding as for instance found in covalent crystals, which are typically brittle. In turn, a positive Cauchy pressure indicates a more metallic bonding as for instance found in transition and noble metals, which are typically ductile.

Phenomenological criteria for intrinsic ductility, such as Refs.~\cite{Kelly:1967,Rice:1974,Rice:1992}, often originate from fracture theories and are expected to give a more physically based insight into intrinsic ductility. 
A commonly employed phenomenological criterion is the one by Rice~\cite{Rice:1992}. It assumes the presence of a sharp, semi-infinite crack in a homogeneous, defect-free material loaded externally. If the crack tip emits a dislocation (thereby blunting the crack) at stress-intensity levels lower than those causing crack cleavage, a material is said to be intrinsically ductile (and vice versa for intrinsically brittle materials).
Rice showed that the criterion for the emission of a dislocation can be associated with the USF energy, whereas that for crack cleavage involves the surface energy. An increment of $\gamma_{\text{s}}/\gamma_{\text{u}}$ is interpreted as a larger likelihood to release stress around a crack tip by slipping of atomic layers, whereas a decrease would crack the material by opening new micro surfaces.

A qualitative connection between Pugh's and Rice's criteria can be established as follows. On the one hand, the surface energy is to a large degree proportional to the bulk modulus since both parameters scale with the cohesive energy as suggested by simple bond-cutting models and corroborated by DFT calculations for transition metals~\cite{kwon2005surface,alden1992calculated,vitos1994full,Moruzzi:1977}.
On the other hand, the USF energy is the energy barrier due to shifting one half of the crystal over the other half, which scales with the shear modulus $G$~\cite{Ogata:2004,li2016generalized}. 
Thus, the trend of the rate of change of $\gamma_{s}/\gamma_{u}$ is expected to be similar to that of $B/G$, at least on a qualitative level.
In this work, we employ four criteria to characterize the ductility of the three considered RAFM steels in comparison to pure Fe: the Poisson ratio $\nu$, the Cauchy pressure $C_{12} - C_{44}$, the Pugh ratio $B/G$, and the Rice ratio $\gamma_{\text{s} (110)}/\gamma_{\text{u} (110)}$ for the usual  $\{110\}\langle 111\rangle$ slip system in the bcc lattice.

The results for $\nu$ and $B/G$ are given in Table~\ref{table:3}, whereas $C_{12} - C_{44}$ is shown in Table~\ref{table:2}, and $\gamma_{\text{s} (110)}/\gamma_{\text{u} (110)}$ is listed in Table~\ref{table:5}.
For Fe$_{91}$Cr$_{9}$, $\gamma_{\text{s} (110)}/\gamma_{\text{u} (110)} = $ 2.204 is only marginally larger than $2.194$, i.e., the averaged ratio for the three RAFM grades.
It should be noted that $\nu$, $C_{12} - C_{44}$, and $B/G$ in Fe$_{100-x}$Cr$_{x}$ for $x\le 9\,$at.\% are all smaller than in Fe (Fig.~\ref{fig:3}), which is consistent with the trend found for the Rice ratio.
Thus, alloying Fe with Cr tends to reduce the intrinsic ductility of Fe.

The decreases of both $B/G$ and Poisson's ratio indicate that CLAM/CLF-1, F82H, and EUROFER97 are intrinsically less ductile than pure Fe. The obtained Cauchy pressure is positive for pure Fe and decreases for the considered RAFM steels, further suggesting that these alloying additions reduce the ductility of Fe.
The computed $\gamma_{\text{s} (110)}/\gamma_{\text{u} (110)}$ ratios for CLAM/CLF-1, F82H, and EUROFER97 are all smaller than that of pure Fe, indicating that the three RAFM steels become more brittle in agreement with the empirical relationships.
Although both $\gamma_{\text{s} (110)}$ and $\gamma_{\text{u} (110)}$ increase with respect to Fe, the alloying effect on $\gamma_{\text{u} (110)}$ is more pronounced, which leads to the negative slope of their mutual ratio.

As in the previous sections, we have a closer look at the Fe$_{91}$Cr$_{9}$ host and investigate how the individual solute elements W, V, and Mn, as well as Cr beyond 9\,at.\% affect the phenomenological ductility indicators. In Fig.~\ref{fig:3}, we show Pugh's ratio, Poisson's ratio, and  Cauchy pressure for Fe$_{91-c}$Cr$_{9}M_c$ with the $c$-ranges and solutes $M$ as before.
For W, we find a strong increase of all three parameters arising primarily from the pronounced positive slopes of $C_{11}$ and $C_{12}$ compared to $C_{44}$.
The effects of V, Mn, and Cr are all substantially smaller. We find a slightly positive slope for V and negative slopes for Mn and Cr for all three criteria.
Thus, the employed phenomenological indicators predict a pronounced and slightly enhanced intrinsic ductility for  W  and V, respectively, whereas Cr or Mn  deteriorate it.

\section{\label{sec:conclusion}Conclusion}

We presented a theoretical investigation of the fundamental
mechanical properties of three reduced activation
ferritic/martensitic steels: CLAM/CLF-1, F82H, and EUROFER97. The
RAFM steels were modeled as quinary FeCrWVMn solid solutions in bcc
phase by means of the exact muffin-tin orbitals method in
combination with the coherent-potential approximation.

The assessment for ferromagnetic bcc Fe and Fe$_{91}$Cr$_{9}$,
selected for its Cr content typical of RAFM steels, showed that the
present theoretical tool captures the compositional trends on the
lattice parameter, single-crystal elastic constants, and
polycrystalline moduli and the predicted alloying effects are in
good agreement with available experimental data. To gain insight
into the mechanical properties of the selected RAFM steels, we chose
the Fe$_{91}$Cr$_{9}$ host as a reference and presented a detailed
analysis of the effects of adding up to 2\,at.\% Mn, 5\,at.\% W or V
(balance Fe), as well as those arising from a 2\,at.\% increase in
the Cr amount. For each solute atom, the total rates of change in
the lattice parameter, shear elastic constants $C^\prime$ and
$C_{44}$, and bulk modulus were decomposed into volumetric and
electronic contributions and analyzed. 
A linear superposition of
these individual rates of change was proposed and validated for the single-crystal and polycrystalline elastic parameters
of the quinary CLAM/CLF-1, F82H, and EUROFER97 model compositions, as well as a
test alloy with higher amounts of Cr, W, V, and Mn.
The data validation showed that the linear description is adequate if the accumulated concentrations of Cr, W, V, and Mn (Cr content relative to 9\, at.\%) is less or equal than 2.4\,at.\%.

The effect of the alloying elements on the intrinsic ductility was
evaluated in the framework of Rice's phenomenological theory.
Accordingly, a competition between Griffith cleavage and dislocation
nucleation is expressed by the ratio of surface energy to unstable
stacking fault energy. We determined both micro-mechanical
parameters for \{110\} planar faults for Fe, Fe$_{91}$Cr$_{9}$,
CLAM/CLF-1, F82H, and EUROFER97, and found close agreement with the
predictions based on empirical relationships employing
single-crystal or polycrystalline elastic parameters.

The present theoretical results provide a comprehensive picture
behind the observed changes in the micro-mechanical properties of
several RAFM steels, and give ground for further theoretical and
experimental efforts for optimizing the composition of these alloys and improve their properties.
Particularly useful for the design of the ferritic phase of RAFM steels is the proposed formula for predicting the lattice parameter, single-crystal elastic constants, and polycrystalline elastic moduli of quinary Fe$_{91-\bar{x}-y-z-u}$Cr$_{9+\bar{x}}$W$_{y}$V$_{z}$Mn$_{u}$ solid solutions from the values computed for the Fe$_{91}$Cr$_{9}$ binary. 

\section{Acknowledgments}

This work was supported by National Magnetic Confinement Fusion
Energy Research Project (2015GB118001), the Swedish Research
Council, the Swedish Foundation for Strategic Research, and the
China Scholarship Council. The Hungarian Scientific Research Fund
(OTKA 109570) are also acknowledged for financial support. The
computations were performed on resources provided by the Swedish
National Infrastructure for Computing (SNIC) at the National
Supercomputer Centre in Link\"oping.

\end{document}